\documentclass[preprint,groupdaddress,aps,pre,amsmath,amssymb,a4paper]{revtex4-1}
 \usepackage[greek,english]{babel}
 \usepackage[iso-8859-7]{inputenc}
 \usepackage[dvips]{graphicx}
 \usepackage[dvips]{graphics}
 \usepackage{epsfig}
  \usepackage{enumerate}
  \usepackage{mathrsfs}
\usepackage{amsmath}
\usepackage{amsfonts}
\usepackage{amssymb}
\usepackage{amsthm}
\usepackage{bm}
\usepackage{graphics}
\usepackage{subfig}
\usepackage{makeidx}
\usepackage{color}
\usepackage{soul}
\usepackage{ragged2e}
\usepackage{fancyhdr}
\usepackage[toc,page]{appendix}
\usepackage{epstopdf}

 \newcommand{\newc}{\newcommand}
 \newc{\ra}{\rightarrow}
 \newc{\lra}{\leftrightarrow}
 \newc{\beq}{\begin{equation}}
 \newc{\eeq}{\end{equation}}
 \newc{\bea}{\begin{eqnarray}}
 \newc{\barra}{\begin{eqnarray*}}
 \newc{\eea}{\end{eqnarray}}
 \newc{\earra}{\end{eqnarray*}}
 \newc{\texa}{\textstyle}
 \newc{\paral}{\parallel}
 \newc{\und}{\underline}
 \newc{\pars}{\partial}
 \newc{\nn}{\nonumber \\}
 \newc{\nln}{\\ \vspace{2mm}}
 \newc{\HRule}{\rule{\linewidth}{0.5mm}}
 \begin{document}
 \title{Symmetry transformations for magnetohydrodynamics and Chew-Goldberger-Low equilibria revisited}
 \author{A. Evangelias}
 \email{aevag@cc.uoi.gr}
\affiliation{University of Ioannina, Department of Physics, Section of Astrogeophysics,
GR 451 10 Ioannina, Greece}

 \author{G. N. Throumoulopoulos}
 \email{gthroum@uoi.gr}
 \affiliation{University of Ioannina, Department of Physics, Section of Astrogeophysics,
GR 451 10 Ioannina, Greece}

\begin{abstract}
Being motivated by the paper [O. I. Bogoyavlenskij, Phys. Rev. E \textbf{66}, 056410 (2002)] we generalise the symmetry transformations for  MHD equilibria with isotropic pressure and  incompressible flow parallel to the magnetic field introduced therein in the case of respective CGL equilibria with anisotropic pressure. We find that the geometrical symmetry of the field-aligned equilibria can break by those transformations  only when the magnetic field is purely poloidal. In this situation we derive three-dimensional CGL equilibria from  given axisymmetric ones. Also, we examine the generic symmetry transformations for MHD and CGL equilibria with incompressible flow of arbitrary direction, introduced in a number of papers, and find that they cannot break the geometrical symmetries of the original equilibria, unless the velocity and magnetic field are collinear and purely poloidal.
\end{abstract}

\maketitle

 \section{\label{1}Introduction}
 
 Two of the most important models widely applied to describe plasma equilibria are the isotropic ideal magnetohydrodynamics (MHD) and the anisotropic Chew-Goldberger-Low (CGL) \cite{CGL} model.
 In \cite{bogo1,bogo,bogo3,bogo2,cheviakov1,cheviakov3,cheviakov2,anco} methods for constructing new continuous families of equilibria in the framework of the above mentioned models, once a given equilibrium is known, are introduced.  More specifically, in \cite{bogo1,bogo,bogo3,bogo2} three sets of equilibrium transformations in the framework of MHD model were presented. The first set is applied to given equilibria with incompressible flow of arbitrary direction, while the second one to both static equilibria and stationary equilibria with field-aligned incompressible flow. The third set of transformations  concerns plasma  equilibria with compressible flow. In addition, in  \cite{cheviakov1} symmetry transformations that produce an infinite class of anisotropic CGL equilibria, on the basis of prescribed CGL ones are introduced; also in \cite{cheviakov1,cheviakov3,cheviakov2,anco} are presented symmetric transformations mapping static or stationary MHD equilibria into CGL ones. All these symmetry transformations depend on a number of scalar functions which have to be constant on the magnetic field  lines. This implies that the new equilibria resulting from the transformations depend on the structure of the magnetic fields of the original ones, and thus, the  topology of the original equilibria is essential for these transformations.
 \par A magnetic field line before closing to itself may either cover a surface, if such a surface exists, or fill a volume.  Any surface that is traced out by a number of magnetic field lines is called magnetic surface. In plasmas of fusion devices, however, the name is usually reserved for nested toroidal surfaces. 
  The generic structure of the magnetic field can be either ``open", in the sense that it closes to itself through infinity, as for example in magnetic mirror, screw pinch and earth's magnetosphere, or closed if it remains in a spatially finite region, as for example in the central region of tokamak and stellarator. 
It can be proved that  if the magnetic field lines  lie on some closed surfaces contained in a bounded  region and do not have any singularities, then they must be toroids (topological tori) \cite{kruskal,hopf,morozov}.  Hamiltonian theory guarantees the existence of magnetic surfaces in systems  with three kinds of continuous geometrical symmetry: axisymmetry, as in an ideal tokamak, helical symmetry which can approximately  describe a ``straight stellarator" without toroidal curvature,  and translational symmetry in which the system is unbounded along the symmetry direction.  However, the latter category can represent a ``straight tokamak" when the magnetic field is periodic along the direction of symmetry and therefore can be considered as toroidal field, since a single period   of such a
field  is topologically equivalent  to a torus.  For such symmetric systems the magnetic surfaces are well-defined by the level sets of a  function $\psi(q^1,q^2)$, with the third coordinate $q^3$  being ignorable. In non-symmetric devices, on the other hand, magnetic surfaces do not exist rigorously everywhere because the magnetic field may cover  regions of finite volume. Open-ended systems, such as magnetic mirrors, do not possess magnetic surfaces that are traced out by one single line. This leads to a considerable degree of arbitrariness.
 \par The lines of force lying on nested  toroidal magnetic surfaces encircle the magnetic axis. This encirclement is characterized by the rotational transform which is defined as the ratio of the number of poloidal transits (the short way around the toroid) to the number of toroidal transits (the long way around it) of a field line. If the rotational transform is a rational number then the magnetic field lines close upon themselves on surfaces that are called rational surfaces,  leaving finite parts of them with vanishing magnetic field.  If not, the surfaces are ergodic (or irrational) and the field lines cover them densely everywhere. In systems without geometrical symmetry there might exist stochastic regions in which the magnetic field lines do not lie on any surfaces but are chaotic, e.g.  near a separatrix. Such regions are undesirable for MHD equilibrium and stability. If the symmetry of the field is violated, for example by superimposing a perturbation having a different symmetry, then an analogous function to $\psi$ may not be found, and magnetic surfaces will no longer be uniquely defined or defined at all \cite{grad}. In \cite{moffatt} it was proved that all smooth steady MHD equilibria with field-aligned incompressible flows possess (open) magnetic surfaces, with possible exception the force-free or Beltrami equilibria. Also, in \cite{moawad} it was proved the existence of (open) magnetic surfaces of three-dimensional equilibria with field-aligned flows.
 \par In \cite{bogo1,bogo,bogo3,bogo2,cheviakov1,cheviakov3,cheviakov2,anco} it is claimed that the symmetry transformations presented therein can break the geometrical symmetries of the original equilibria either static or with parallel incompressible flows. In this context, the respective symmetry transformations were applied to both the magnetic analog of Hill's vortex static axisymmetric equilibria with purely poloidal magnetic field \cite{shafranov, hill}, and to static helically symmetric equilibria with magnetic field along the symmetry direction \cite{bogo2,johnson} in order to model nonsymmetric astrophysical jets. 

In the present work we make an extensive revision of the transformations presented previously in Refs. \cite{bogo1,bogo,bogo3,bogo2,cheviakov1,cheviakov3,cheviakov2,anco} concerning equilibria with incompressible flows. In Section \ref{2} we introduce a symmetry transformation that can be applied to any known anisotropic CGL equilibria with field-aligned incompressible flows (or static equilibria) and anisotropy function constant on magnetic field lines, and produce an  infinite family of anisotropic equilibria with collinear velocity and magnetic fields, but density and anisotropy functions that may remain arbitrary. These transformations consist a generalisation of the ones introduced in \cite{bogo1} for field-aligned MHD equilibria. We also prove that all transformations presented  in \cite{bogo1,bogo,bogo3,bogo2,cheviakov1,cheviakov3,cheviakov2,anco} can break the geometrical symmetries of a known given equilibrium, static or with field-aligned flow,  if and only if its magnetic field is purely poloidal. In Section \ref{3} we construct three-dimensional (3D) equilibria by applying the introduced transformations to known axisymmetric equilibria with field-aligned incompressible flow, pressure anisotropy, and purely poloidal magnetic field, related with the symmetry breaking. In Section \ref{4} we examine the aforementioned symmetry transformations for flow of arbitrary direction and check their validity in connection with  the structure of the magnetic fields of the original equilibria and the existence of magnetic surfaces. Finally, Section \ref{5} summarizes the conclusions.
 \section{\label{2}Symmetry transformations for field-aligned equilibria}
 \subsection{\label{2.1}Review of the transformations for MHD equilibria}
 In Section IV of Ref. \cite{bogo1} transformations between MHD equilibria with parallel flows are presented. Specifically, it is stated therein that if $\{ \vec{B},\, \vec{v},\, p,\, \varrho \}$ is a solution of the ideal MHD equilirium system of equations with field-aligned incompressible flow:
 \begin{eqnarray}
 \begin{gathered}
 \label{MHDpar}
\varrho(\vec{v}\cdot\vec{\nabla})\vec{v}=\vec{J}\times\vec{B}-\vec{\nabla}p, \quad \vec{\nabla}\cdot\vec{B}=0,\\
\vec{\nabla}\cdot (\varrho\vec{v})=0, \quad \vec{\nabla}\times\vec{B}=\mu_{0}\vec{J},
 \end{gathered}
 \end{eqnarray}
  then $\{ \vec{B}_1,\, \vec{v}_1,\, p_1,\, \varrho _1 \}$ defined by the following symmetry transformations, that depend on the arbitrary functions $a(\vec{r}),\, b(\vec{r}),\, c(\vec{r})$,  consist a new solution to the MHD equilibrium set of equations with field-aligned flows:
  \begin{eqnarray}
 \begin{gathered}
 \label{transfMHDpar}
\vec{B}_1=b(\vec{r})\vec{B}, \quad \vec{v}_1=\frac{c(\vec{r})}{a(\vec{r})\sqrt{\mu _0 \varrho}}\vec{B},\\
\varrho _1 (\vec{r})=a^2(\vec{r})\varrho , \quad p_1=C\left(p+\frac{B^2}{2\mu _0}\right)-\frac{B_1^2}{2\mu _0},\\
 C= \frac{b^2(\vec{r})-c^2(\vec{r})}{1-\lambda ^2(\vec{r})}=\mbox{const.}\neq 0.
\end{gathered}
\end{eqnarray}
 The above special transformations are defined only when the velocity and magnetic field of the original equilibria are related through $ \vec{v}=(\lambda/\sqrt{\mu _0 \varrho})\vec{B}$, and are also valid in the static limit, $\vec{v}=0$. Their reductive form for constant $a, \, b, \, c$, and $\lambda$ was first derived in \cite{bogo3} from given axisymmetric equilibria found in \cite{bogo4}.
According to \cite{bogo1} the functions $a(\vec{r})$, $b(\vec{r})$,
 $c(\vec{r})$, depending on the topology of the original equilibria may either (i) be constant on magnetic surfaces, or (ii) in case of symmetry involving  two dimensional dependence, depend on  two transversal variables (i.e. variables not dependent explicitly on the ignorable coordinate), or (iii) be constants in the case of force-free equilibria. Also it is claimed therein that transformations (\ref{transfMHDpar}) can break the geometrical symmetry of the original equilibria (\ref{MHDpar}) with general field-aligned incompressible flow.
 \subsection{\label{2.2}Generalised symmetry transformations for anisotropic pressure}
 \par
 In the present  Subsection  we first generalise the transformations (\ref{transfMHDpar}) introduced in \cite{bogo1} for CGL anisotropic equilibria with field-aligned incompressible flow and show that the only situation in which the symmetry of the original equiliria can break is that for purely poloidal magnetic fields. These considerations are summarized in the following theorem:
 \newline
 \newline
 \textbf{Theorem 1.}
 Let $\{\vec{B},\, \vec{v},\, \varrho ,\, p_{\perp},\, p_{\parallel}  \}$ be a known solution to the CGL equilibrium system of equations with field-aligned incompressible flows,
 \begin{equation}
 \label{fieldaligned1}
 \vec{v}=\frac{\lambda (\vec{r})}{\sqrt{\mu _0 \varrho}}\vec{B},
 \end{equation}
 and pressure anisotropy function, 
 \begin{equation}
 \sigma _{d}:=\mu _0 \frac{p_{\parallel}-p_{\perp}}{B^2}, 
  \end{equation}
 constant on magnetic field lines. Then $\{\vec{B}_1,\, \vec{v}_1, \, \varrho _1,\, p_{\perp _1},\, p_{\parallel _1}  \}$
  given by the following transformations:
  \begin{eqnarray}
 \begin{gathered}
 \label{transfCGLpar}
  \vec{B}_1=\frac{b(\vec{r})}{n(\vec{r})}\vec{B},\quad \vec{v}_1=\frac{c(\vec{r})\sqrt{1-\sigma _d}}{a(\vec{r})\sqrt{\mu _0 \varrho}}\vec{B},\\
 \varrho _1 (\vec{r})=a^2(\vec{r})\varrho ,\quad p_{\perp 1}=C\left(p_{\perp}+\frac{B^2}{2\mu _0}\right)-\frac{B_1^2}{2\mu _0}, \\
 p_{\parallel 1}=C\left(p_{\perp}+\frac{B^2}{2\mu _0}\right)+\left[1-2n^2(\vec{r})(1-\sigma _d)\right]\frac{B_1^2}{2\mu _0}, \\
 C= \frac{\left[b^2(\vec{r})-c^2(\vec{r})\right](1-\sigma _d)}{1-\sigma _d-\lambda ^2(\vec{r})}=\mbox{const.}\neq 0,
 \end{gathered}
 \end{eqnarray}
 where $a(\vec{r})\neq 0,\, b(\vec{r}), \, c(\vec{r})$, and $n(\vec{r})\neq 0$, are arbitrary functions, define a solution to the CGL set of equilibrium equations with field-aligned flows, if and only if the functions 
 \begin{equation}
 \label{gf}
 g(\vec{r}):= \frac{b(\vec{r})}{n(\vec{r})}, \quad f(\vec{r}):= a(\vec{r})c(\vec{r}),
 \end{equation}
  are constant on the magnetic field lines of the original equilibria. 
 \newline
 \newline
 \textbf{Proof.}
 The original equilibria $\{\vec{B},\, \vec{v},\, \varrho ,\, p_{\perp},\, p_{\parallel}  \}$ satisfy the CGL equilibrium equations with field-aligned flows (\ref{fieldaligned1}):
   \begin{eqnarray}
   \begin{gathered}
   \label{CGLpar}
   \varrho(\vec{v}\cdot\vec{\nabla})\vec{v}=\vec{J}\times\vec{B}-\vec{\nabla}\cdot{\bf{P}}, \quad \vec{\nabla}\cdot\vec{B}=0, \\
   \vec{\nabla}\cdot (\varrho\vec{v})=0, \quad \vec{\nabla}\times\vec{B}=\mu_{0}\vec{J},
\end{gathered}
\end{eqnarray}
where the CGL pressure tensor is defined as
\begin{equation}
{\bf P}:=p_{\perp}{\bf I}+\frac{\sigma_{d}}{\mu_{0}}\vec{B}\vec{B},
\end{equation}
with the function $\sigma _d$ measuring the pressure anisotropy. It is assumed that the flow is incompressible, $\vec{\nabla}\cdot \vec{v}=0$, which by the continuity equation
 implies that the mass density is constant on streamlines, $\vec{v}\cdot \vec{\nabla}\varrho (\vec{r})=0$; it is also assumed  that the anisotropy function is constant on the magnetic field lines, $\vec{B}\cdot \vec{\nabla}\sigma _d (\vec{r})=0$. When the equilibria possess some geometrical symmetry, the latter hypothesis for the function $\sigma _d$ in conjunction  with incompressibility, lead to the derivation of a single Grad-Shafranov (GS) equation that governs them \cite{clemente,evangelias1,evangelias2}; also, according to \cite{mercier} this assumption on $\sigma_d$ may be the only suitable for satisfying the boundary conditions on a fixed, perfectly conducting wall. It may be noted that for the given field-aligned equilibria, the vectors $\vec{v}$ and $\vec{B}$ are collinear (parallel) and therefore  the magnetic field lines are the same as the  velocity streamlines. It follows that the function $\lambda (\vec{r})$ must be constant on magnetic field lines,  $\vec{B}\cdot \vec{\nabla}\lambda (\vec{r})=0$.
Also,  the force balance equation of the set (\ref{CGLpar}) can be cast into the useful form
\begin{equation}
\label{forcebalance}
(1-\sigma _d-\lambda ^2)\frac{1}{\mu _0}\vec{B}\times(\vec{\nabla}\times \vec{B})+(\sigma _d+\lambda ^2)\vec{\nabla}\left(\frac{B^2}{2\mu _0}\right)+\vec{\nabla}p_{\perp}=0.
\end{equation}
In order for the new solution (\ref{transfCGLpar}) to be valid it must satisfy the following set of CGL equilibrium equations:
\begin{eqnarray}
\begin{gathered}
\label{CGLparnew}
\varrho _1(\vec{v}_1\cdot\vec{\nabla})\vec{v}_1=\vec{J}_1\times\vec{B}_1-\vec{\nabla}\cdot{\bf P}_1, \quad \vec{\nabla}\cdot\vec{B}_1=0, \\
\vec{\nabla}\cdot (\varrho _1\vec{v}_1)=0, \quad \vec{\nabla}\times\vec{B}_1=\mu_{0}\vec{J}_1,
\end{gathered}
\end{eqnarray}
where
\begin{eqnarray} 
\label{tensor}
{\bf P}_1:=p_{\perp _1}{\bf I}+\frac{\sigma_{d_1}}{\mu_{0}}\vec{B}_1\vec{B}_1 , \quad \sigma_{d_1}:= \mu _0 \frac{p_{\parallel _1}-p_{\perp _1}}{B_1^2}=1-n^2(\vec{r})(1-\sigma _d).
\end{eqnarray}
Note that systems (\ref{CGLpar})  and  (\ref{CGLparnew}) are   reductions of the generic CGL equilibrium equations since for field-aligned flows it holds $\vec{v}\times \vec{B}=\vec{v}_1\times \vec{B}_1=0$, and therefore the electric field vanishes by Ohm's law.
\par
Substituting (\ref{transfCGLpar}) into (\ref{CGLparnew}) yields
\begin{eqnarray}
\label{first}
\vec{B}\cdot \vec{\nabla}\left(\frac{b(\vec{r})}{n(\vec{r})}\right)=0, \\
\label{second}
\vec{B}\cdot \vec{\nabla}(a(\vec{r})c(\vec{r}))=0, \\
\label{third}
C\left[(1-\sigma _d-\lambda ^2)\frac{1}{\mu _0}\vec{B}\times(\vec{\nabla}\times \vec{B})+(\sigma _d+\lambda ^2)\vec{\nabla}\left(\frac{B^2}{2\mu _0}\right)+\vec{\nabla}p_{\perp}\right]\nonumber \\
-\frac{1-\sigma _d}{\mu _0}\vec{B}\cdot 
\left(b^2\frac{\vec{\nabla}n}{n}+c^2\frac{\vec{\nabla}a}{a}\right)=0.
\end{eqnarray}
With the use of Eq. (\ref{forcebalance}) and assuming that $\sigma _d\neq 1$ (in which case $\vec{v}_1=0$, $C=0$ and the transformations
(\ref{transfCGLpar}) are not invertible),  Eq. (\ref{third}) takes the form
\begin{eqnarray}
\label{validationCGLpar3}
\vec{B}\cdot \left(b^2\frac{\vec{\nabla}n}{n}+c^2\frac{\vec{\nabla}a}{a}\right)=0.
\end{eqnarray}
Now with the aid of (\ref{gf}),
Eqs. (\ref{first}), (\ref{second}) and (\ref{validationCGLpar3}) assume the forms
\begin{eqnarray}
\label{first1}
\vec{B}\cdot \vec{\nabla}g(\vec{r})=0, \\
\label{second1}
\vec{B}\cdot \vec{\nabla}f(\vec{r})=0, \\
\label{third1}
\frac{C}{2}\underbrace{\vec{B}\cdot \vec{\nabla}\left(\frac{1-\sigma _d-\lambda}{1-\sigma _d}\right)}_{\text{0}}+\frac{c^2(\vec{r})}{f(\vec{r})}\vec{B}\cdot \vec{\nabla}f(\vec{r})-\frac{b^2(\vec{r})}{g(\vec{r})}\vec{\nabla}g(\vec{r})=0.
\end{eqnarray}
Since the first term on the lhs of (\ref{third1}) vanishes, it is apparent  that if  Eqs.  (\ref{first1}) and (\ref{second1}) are valid, then (\ref{third1}) is trivially satisfied. Thus, we conclude that in order for transformations (\ref{transfCGLpar}) to be valid, Eqs.  (\ref{first1}) and (\ref{second1}) must be satisfied, or equivalently, both functions $g(\vec{r})$ and $f(\vec{r})$ have to be constant on the magnetic field lines of the original equilibria; QED
\newline
 \newline
 \textbf{Remark 1.}
 The symmetry transformations (\ref{transfCGLpar}) presented herein are defined only when the velocity and the magnetic field of the original equilibria are related through (\ref{fieldaligned1}), and transformations (\ref{transfMHDpar}) introduced in \cite{bogo1} for field-aligned MHD equilibria consist a special case of them for $\sigma _d=0$ and $n(\vec{r})=1$.
 \newline
\par Let us now examine the structure of the arbitrary scalar functions in connection with the magnetic field by first noting that the magnetic and velocity fields of the original equilibria (\ref{fieldaligned1}) lie on the surfaces $\lambda (\vec{r}) =\mbox{const.}$, $
\vec{B}\cdot \vec{\nabla}\lambda (\vec{r})=\vec{v}\cdot \vec{\nabla}\lambda (\vec{r})=0$,
if such surfaces exist.
\par
If the magnetic field lines are spatially bounded closed curves, labeled  by  $l$, then the surfaces $\lambda =\mbox{const.}$ are defined in the neighbourhood of $l$, with $l$  being  itself both a magnetic field line and a streamline. In this situation, the function $\lambda$ depends on two transversal variables which must define a plane normal to every point of $l$.
If the magnetic field lines are ``open'', i.e. they approach infinity in one direction, defined by a variable $q^3$, then the magnetic field should be finite as $q^3\rightarrow \infty$. The function $\lambda(\vec{r})$ should depend on two transversal variables when the third one goes to infinity, $\lambda(q^1,q^2,q^3\rightarrow \infty)=\lambda(q^1,q^2)$, and thus, must depend on these two variables in the whole plasma domain for magnetic surfaces $\lambda(q^1,q^2)=\mbox{const.}$ to exist. 
 In both of the above kinds of magnetic field lines (bounded and ``open''), the functions  $g(\vec{r})$, $f(\vec{r})$ have to be functions of two transversal variables, i.e. $q^1,q^2$. One could suggest that this does not restrict the functions $a(\vec{r})$, $b(\vec{r})$, $c(\vec{r})$, $n(\vec{r})$ to have the same two-dimensional dependency (i.e. $a(\vec{r})=A(q^1,q^2)D(q^3)$ and $c(\vec{r})=K(q^1,q^2)/D(q^3)$, such that $f(\vec{r})=A(q^1,q^2)K(q^1,q^2)$). However, equation $\vec{\nabla}\cdot(\varrho _1 \vec{v_1})=0$ yields
$\vec{B}\cdot \vec{\nabla}a(\vec{r})=0$, 
which means that $a=a(q^1,q^2)$ and consequently $c=c(q^1,q^2)$. Then from the definition of the constant $C$ it follows that  $b=b(q^1,q^2)$, and as a result $n=n(q^1,q^2)$. Thus, if the magnetic field lines are finite closed loops or go to infinity in some direction, all functions of transformations must, in general, depend on two variables transversal to this direction.
\par If the magnetic field lines cover densely everywhere (ergodically) closed magnetic surfaces, $\lambda(\vec{r})=\mbox{const.}$ (which are toroids), then the functions $g(\vec{r})$, $f(\vec{r})$ must be constant on them, and so must be all four functions of the transformations. In this situation, if the field possesses some geometrical symmetry, with ignorable variable $q^3$, the surfaces $\lambda(\vec{r})=\mbox{const.}$ are nested, with $\lambda=\lambda (q^1,q^2)$. Then all functions $a(\vec{r})$, $b(\vec{r})$, $c(\vec{r})$, $n(\vec{r})$ have the same symmetry (i.e. are functions only of $q^1,q^2$). However, there exists an exception; the one when the original equilibrium has some known geometrical symmetry with purely poloidal magnetic field to be examined as follows.
\par
{\em Axial symmetry:} Consider the case that the original equilibria are axially symmetric with field-aligned incompressible flows and anisotropy function constant on magnetic surfaces \cite{evangelias1}. Employing cylindrical coordinates $(\rho, \, z, \, \phi)$ we have
\begin{equation}
\vec{B}=\frac{I}{\rho}\hat{\phi}+\frac{\hat{\phi}}{\rho}\times \vec{\nabla}\psi(\rho ,z) , \quad
\vec{v}=\frac{M_p}{\sqrt{\mu _0 \varrho}}\vec{B},
\end{equation} 
where the function $I$ relates to the toroidal magnetic field and $\psi(\rho ,z)=\mbox{const.}$ labels the magnetic surfaces. Thus, $\lambda(\vec{r})=\lambda(\psi)=M_p(\psi)$, where $M_p:=(\sqrt{\mu _0 \varrho}v_{pol})/B_{pol}$ is the poloidal Alfv\' en Mach function, which for parallel flows equals to the total Mach function ($M=\sqrt{\mu _0 \varrho}v/B$). To examine whether transformations (\ref{transfCGLpar}) can break axisymmetry we permit the transformation functions to depend, in addition to $\psi$, explicitly  on $\phi$, i.e. $f=f(\psi,\phi)$, $g=g(\psi,\phi)$. Then  (\ref{first1})-(\ref{second1}) yield
\begin{eqnarray}
\begin{gathered}
\label{axial}
\frac{I}{\rho ^2}\left(\frac{\partial g}{\partial \phi}\right)=0, \\
\frac{I}{\rho ^2}\left(\frac{\partial f}{\partial \phi}\right)=0 .
\end{gathered}
\end{eqnarray}
Set (\ref{axial}) is satisfied either if functions $g,f$ are constant on the magnetic surfaces, or $I=0$. The latter case implies that transformations (\ref{transfCGLpar}) can break the axial symmetry of field-aligned equilibria with purely poloidal magnetic field. 
The same statement holds for translationally symmetric equilibria with field-aligned flows \cite{throumtasso}, while the more generic case of helical symmetry will be studied separately below.
\par
{\em Helical symmetry:} Consider now that the original equilibria are helically symmetric with field-aligned  incompressible flows and anisotropy function constant on magnetic surfaces for which the following relations  hold \cite{evangelias2}:
\begin{equation}
\label{Bhel}
\vec{B}=I\vec{h}+\vec{h}\times \vec{\nabla}\psi(r,u), \quad \vec{v}=\frac{M_p(\psi)}{\sqrt{\mu _0 \varrho}}\vec{B}.
\end{equation}
Here $(r,u,\xi)$ are helical coordinates defined through the usual cylindrical ones $(\rho,\phi,z)$ as $r=\rho$, $u=m\phi -kz$, $\xi=z$; $I$ relates to the helicoidal magnetic field and $\psi(r,u)=\mbox{const.}$ labels the magnetic surfaces; the vector $\vec{h}=(m/(k^2r^2+m^2))\vec{g}_{\xi}$ points along the helical direction, where $(k, \, m)$  are integers,   and the covariant helical basis vectors, $\vec{g}_i\, ,i=r,u,\xi $,  are defined through the respective cylindrical unit vectors as $\vec{g}_{r}=\hat{\rho}$, $\vec{g}_{u}=(r/m)\hat{\phi}$, $\vec{g}_{\xi}=(rk/m)\hat{\phi}+\hat{z}$.
For the adopted non-orthogonal helical coordinates, the $u$-  and $\xi $-covariant and contravariant components of a given vector $\vec{A}$ differ from  each other, $A_i\neq A^i \ ( i=u, \xi$).
The magnetic field written in contravariant components is
\begin{equation}
\label{Bcontra}
\vec{B}=B^{\xi}\vec{g}_{\xi}+B^r\vec{g}_r+B^u\vec{g}_u.
\end{equation}
The definition of a usual flux function $\psi (r,u)$ so as equation $\vec{\nabla}\cdot \vec{B}=0$ to be satisfied reduces (\ref{Bcontra}) into [see also \cite{agim}]:
\begin{equation}
\vec{B}=B^{\xi}\vec{g}_{\xi}+\frac{1}{m}\vec{g}^{\xi}\times \vec{\nabla}\psi ,
\end{equation}
where $\vec{g}^{\xi}=\hat{z}$ is the contravariant basis vector.
Observe that 
\begin{equation}
\vec{g}_{\xi}\cdot (\vec{g}^{\xi}\times \vec{\nabla}\psi)=\frac{kr}{m}\frac{\partial \psi}{\partial r}\neq 0.
\end{equation}
 This in fact  dictated us to define the helical vector $\vec{h}$ that points into the symmetry direction. Then the magnetic field is written in the form (\ref{Bhel})  with 
\begin{equation}
I\equiv \frac{B^{\xi}}{mq}+\frac{kr}{m}\frac{\partial \psi}{\partial r}.
\end{equation}
If we define the poloidal magnetic field as
\begin{equation}
\vec{B}_{pol}=\frac{1}{m}\vec{g}^{\xi}\times \vec{\nabla}\psi ,
\end{equation}
then the field on the plane normal to $\vec{h}$ is expressed as
\begin{equation}
 \vec{h}\times \vec{\nabla}\psi =\vec{B}_{pol}-kqr\frac{\partial \psi}{\partial r}\vec{g}_{\xi}.
\end{equation}
Now let $f=f(\psi,\xi)$, $g=g(\psi,\xi)$. Then satisfaction of Eqs.  (\ref{first1})-(\ref{second1}) require
\begin{eqnarray}
\begin{gathered}
\label{helical}
B^{\xi}\left(\frac{\partial g}{\partial \xi}\right)=0, \\
B^{\xi}\left(\frac{\partial f}{\partial \xi}\right)=0.
\end{gathered}
\end{eqnarray}
Respectively to (\ref{axial}), Eq. (\ref{helical}) leads to
\begin{equation}
 B^{\xi}=0 \Rightarrow  I=\frac{kr}{m}\frac{\partial \psi}{\partial r} \Rightarrow \vec{B}=\vec{B}_{pol}.
 \end{equation}
 Thus, transformations (\ref{transfCGLpar}) can also break the helical symmetry of the original equilibrium with field-aligned incompressible flow and pressure anisotropy, if and only if the magnetic field is purely poloidal.
 \newline
 \par Finally, it may happen that $\lambda=\mbox{constant}$ and consequently, $\vec{\nabla}\lambda=0$ in the whole plasma domain. In this  situation the force balance equation (\ref{forcebalance}) is written in the form
 \begin{equation}
 (1-\sigma _d-\lambda ^2)\vec{J}\times \vec{B}=\vec{\nabla}\left(\bar{p}+\lambda ^2\frac{B^2}{2\mu _0}\right)-\frac{B^2}{2\mu _0}\vec{\nabla}\sigma _d ,
 \end{equation}
 where $\bar{p}:=(p_{\perp}+p_{\parallel})/2$ is defined as an effective isotropic pressure.
 In this case a family of magnetic surfaces $w(\vec{r})=\mbox{const.}$, where $w\equiv \bar{p}+\lambda ^2\frac{B^2}{2\mu _0}$ can be defined, in which both magnetic field lines and velocity streamlines lie on, $\vec{B}\cdot \vec{\nabla}w(\vec{r})=0$. Analogous considerations can be made on the structure of these surfaces.
 \par
 Now it may happen $w=\mbox{const.}$ with $\vec{\nabla}w=0$ if $\vec{J}=y(\vec{r})\vec{B}$, that is the current density is parallel to the magnetic field,  or equivalently $\vec{\nabla}\times\vec{v}=t(\vec{r})\vec{v}$, that is the velocity is parallel to the vorticity. This is the case of  force free or Beltrami equilibria. Then magnetic surfaces $y(\vec{r})=\mbox{const.}$ can be yet defined, $\vec{B}\cdot \vec{\nabla}y(\vec{r})=0$. But in the particular case  $y\equiv\mbox{const.}$ (everywhere) and therefore  $\vec{\nabla}y=0$ (and then as well $t\equiv\mbox{const.}$, with $\vec{\nabla}t=0$), we finally escape from the topological constraint that magnetic field lines lie on surfaces. The lines of force may be chaotic (space-filling) in this case, and all functions $a(\vec{r})$, $b(\vec{r})$, $c(\vec{r})$, $n(\vec{r})$ have to be constant.
 \par The above conclusions lead us to formulate the following  corollary:
\newline
\newline
\textbf{Corollary 1.} 
Transformations (\ref{transfCGLpar}) can break the geometrical symmetry, either axial or translational or helical, of the original field-aligned equilibria with incompressible flow and anisotropy function constant on magnetic surfaces, if and only if its magnetic field is purely poloidal. Otherwise, the transformed equilibria retain the original symmetry.
\newline
\par
All conclusions derived herein concerning the validity of the transformations, the structure of the arbitrary functions and the symmetry breaking, also hold for the respective transformations (\ref{transfMHDpar}) with isotropic pressure (cf Remark 1).
 In this respect, the symmetry breaking of the static helically symmetric equilibria related with astrophysical jets, examined in Section VIII of Ref. \cite{bogo1}, should be revised, since the magnetic field of the original equilibria [cf Eq. (8.1) therein] is not purely poloidal unless the constant $\alpha$ is equal to zero.
 \section{\label{3}Construction of 3D CGL equilibria with field-aligned flows}
 Consider axisymmetric equilibria \cite{evangelias1} with field-aligned incompressible flows, pressure anisotropy and purely poloidal magnetic field. In this case the equilibrium quantities are expressed as 
 \begin{eqnarray}
 \begin{gathered}
 \vec{B}=\frac{\hat{\phi}}{\rho}\times \vec{\nabla}\psi , \quad \mu _0 \vec{J}=\frac{1}{\rho}\Delta ^{*}\psi \hat{\phi} \\
 \vec{v}=\frac{M_p}{\sqrt{\mu _0 \varrho}}\vec{B}, \quad
 \bar{p}=\bar{p}_s(\psi)-M_p^2\frac{B^2}{2\mu _0},
 \end{gathered}
 \end{eqnarray}
 and the steady sates  obey the following  generalized GS equation
 \begin{equation}
 \label{GS}
 (1-\sigma_d-M_p^2)\Delta ^{*}\psi+\frac{1}{2}\frac{d(1-\sigma_d-M_p^2)}{d\psi}|\vec{\nabla}\psi |^2+\mu _0 \rho ^2\frac{d\bar{p}_s}{d\psi}=0, \quad \sigma _d +M_p^2<1.
 \end{equation}
 Here the elliptic operator is defined as $\Delta ^{*}:=\rho ^2 \vec{\nabla}\cdot\left(\vec{\nabla}/\rho ^2\right)$; $\bar{p}_s$ is the effective pressure in the absence of flow, and the functions $\varrho , \, \sigma _d, \, M_p$ are constant on the magnetic surfaces $\psi =\mbox{const.}$
 Applying the symmetry transformations (\ref{transfCGLpar}), with $\lambda=M_p(\psi)$, we obtain the following 3D equilibria:
 \begin{eqnarray}
 \begin{gathered}
 \label{3dequil}
 \vec{B}_1=\frac{b}{n}\frac{\hat{\phi}}{\rho}\times \vec{\nabla}\psi ,\quad
 \vec{v}_1=\frac{c}{a}\frac{\sqrt{1-\sigma _d}}{\sqrt{\mu _0 \varrho}}\frac{\hat{\phi}}{\rho}\times \vec{\nabla}\psi ,\\
 \mu _0 \vec{J}_1=\frac{1}{\rho}\left[\frac{b}{n}\Delta ^{*}\psi +\frac{\partial (b/n)}{\partial \psi}|\vec{\nabla}\psi |^2\right]\hat{\phi}-\frac{\partial (b/n)}{\partial \phi}\frac{\vec{\nabla}\psi}{\rho ^2},\\
\varrho _1=a^2\varrho , \quad \bar{p}_1=C\bar{p}_s(\psi)-c^2(1-\sigma _d)\frac{B^2}{2\mu _0},
\end{gathered}
 \end{eqnarray}
 where the functions $a, \, b, \, c,$ and $n$ may depend, in addition to $\psi$, on the toroidal angle $\phi$. However, if either of the functions $n(\vec{r})$ or $a(\vec{r})$ remain constant on magnetic surfaces, the breaking of the geometrical symmetry of the original equilibria remains unaffected.
 Note that the transformed current density $\vec{J}_1$ has a component perpendicular to the magnetic surfaces which is undesirable for confinement but this component vanishes when the function $g=b/n$ is $\phi$-independent. This choice, however, yields special equilibria with purely poloidal magnetic field, $\vec{B}_1=\kappa (\psi) \vec{B}$ and permits  only  3D variations of velocity and pressure.
 \par To construct a specific equilibrium let us make the following choice for the arbitrary functions:
\begin{eqnarray}
\begin{gathered}
c(\psi , \phi)=[cos(2\delta \phi)+A_0]\left(\frac{1-\sigma _d-M_p^2}{1-\sigma _d}\right)^{1/2}, \quad |A_0|>1,\\
  b(\psi , \,\phi)=\sqrt{[cos(2\delta \phi)+A_0]^2+1}\left(\frac{1-\sigma _d-M_p^2}{1-\sigma _d}\right)^{1/2}, \\
  a(\psi)=\frac{1}{2}(1-\sigma _d)^{1/2}, \quad n(\psi ,\,\phi)=a(\psi)b(\psi , \, \phi).
  \end{gathered}
 \end{eqnarray}
 Then from (\ref{3dequil}) we obtain the exact equilibria with purely poloidal magnetic field, incompressible flows, and anisotropy function varying on the magnetic surfaces:
  \begin{eqnarray}
  \begin{gathered}
  \label{new3d}
 \vec{B}_1=\frac{2}{\sqrt{1-\sigma_d}}\vec{B} ,\quad  
\vec{v}_1=[cos(2\delta \phi)+A_0]\sqrt{\frac{1-\sigma _d-M_p^2}{\mu_0 \varrho}}\vec{B}_1 ,\\
 \varrho _1=\frac{(1-\sigma _d)}{4}\varrho ,\quad
 \vec{J}_1=\frac{2}{\sqrt{1-\sigma_d}}\left[\vec{J}+\frac{|\vec{\nabla}\psi |^2}{2(1-\sigma _d)\mu _0 \varrho}\left(\frac{d\sigma _d}{d\psi}\right)\hat{\phi}\right], \\
 \bar{p}_1=\bar{p}_s(\psi)-[cos(2\delta \phi)+A_0]^2(1-\sigma _d-M_p^2)\frac{B^2}{2\mu _0}, \\
 \sigma _{d_1}=1-\left([cos(2\delta \phi)+A_0]^2+1\right)\frac{(1-\sigma _d-M_p^2)(1-\sigma _d)}{4}.
 \end{gathered}
 \end{eqnarray}
 We note that the above equilibria do not obey a GS-like equation analogous to (\ref{GS}). In order for equilibria (\ref{new3d}) to be physically plausible we require $\bar{p}_1>0$, which yields the following restriction for $M_p^2+\sigma_d$:
 \begin{equation}
 1-\frac{\beta _s}{\left(1+A_0\right)^2}<M_p^2+\sigma_d,
 \end{equation}
 where $\beta _s:= \bar{p}_s/(B^2/2\mu _0)$  is the poloidal beta in the absence of flow. It is also recalled that $M_p^2+\sigma_d<1$  [cf  Eq.  (\ref{GS})].
  \par
 If the plasma is confined for example in an axisymmetric device then the physical quantities should be periodic in the angle $\phi$. This yields for $\delta$:
 \begin{equation}
\cos(2\delta \phi)=\cos\left\lbrack2\delta (\phi+2\pi)\right\rbrack\Rightarrow \delta=\frac{l}{2}, \, \, l=1,2,3,... 
 \end{equation}
 Though it is well known that toroidal plasma confinement is not possible with a purely poloidal magnetic field, it is interesting that in that case transformations (\ref{transfCGLpar}) can break the geometrical symmetry and yield 3D equilibria. These equilibria may be of astrophysical interest.
 \section{\label{4}Transformations for flow of arbitrary direction}
 \subsection{\label{4.1}Review of transformations between MHD-MHD and CGL-CGL equilibria}
 In \cite{bogo1,bogo,cheviakov1} symmetry transformations that produce an infinite family of MHD (CGL) equilibria with arbitrary incompressible flow once a respective MHD (CGL) equilibrium with incompressible flow is given, were introduced as follows. 
 
  {\em MHD into MHD:}  In the case of isotropic pressure,  suppose that $\{ \vec{B},\,  \vec{v},\, p,\, \varrho \}$ is a known solution of the MHD equilibrium system with flow of arbitrary direction
 \begin{eqnarray}
 \begin{gathered}
\label{MHD}
\varrho(\vec{v}\cdot\vec{\nabla})\vec{v}=\vec{J}\times\vec{B}-\vec{\nabla}p, \quad \vec{\nabla}\cdot (\varrho\vec{v})=0, \\
\vec{\nabla}\times\vec{B}=\mu_{0}\vec{J},\quad \vec{\nabla}\cdot\vec{B}=0, \quad
\vec{v}\times\vec{B}=\vec{\nabla}\Phi ,
\end{gathered}
\end{eqnarray}
 where $\Phi$ is the electrostatic potential. The flow is assumed to be incompressible, $\varrho = \varrho (\psi)$, and the function $\psi $  labels  the common magnetic and velocity surfaces,  if such surfaces exist. Note that these two sets of surfaces should coincide for flows of arbitrary direction because of the Faraday's and Ohm's laws. Then according to \cite{bogo1,bogo},  $\{ \vec{B}_1,\, \vec{v}_1,\, p_1,\, \varrho _1\}$ defined by the following symmetry transformations (depending  on the arbitrary functions $a(\vec{r}),\, b(\vec{r}),\, c(\vec{r})$)
 \begin{eqnarray}
 \begin{gathered}
 \label{transfMHD}
 \vec{B}_1=b(\vec{r})\vec{B}+c(\vec{r})\sqrt{\mu _0 \varrho}\vec{v},\quad 
 \vec{v}_1=\frac{c(\vec{r})}{a(\vec{r})\sqrt{\mu _0 \varrho}}\vec{B}+\frac{b(\vec{r})}{a(\vec{r})}\vec{v}, \\
 \varrho _1 (\vec{r})=a^2(\vec{r})\varrho ,\quad
 p_1=C\left(p+\frac{B^2}{2\mu _0}\right)-\frac{B_1^2}{2\mu _0}, \\
 C\equiv b^2(\vec{r})-c^2(\vec{r})=\mbox{const.}\neq 0,
 \end{gathered}
 \end{eqnarray}
 consist a new family of solutions to the MHD equilibrium system.
 
{\em CGL into CGL:}  For anisotropic pressure  let $\{\vec{B},\, \vec{v},\, \rho ,\, p_{\perp},\, p_{\parallel}  \}$ be a given solution of the CGL equilibrium system of equations
 \begin{eqnarray}
 \begin{gathered}
\label{CGL}
\varrho(\vec{v}\cdot\vec{\nabla})\vec{v}=\vec{J}\times\vec{B}-\vec{\nabla}\cdot{\bf{P}},\quad \vec{\nabla}\cdot (\varrho\vec{v})=0,
 \\
\vec{\nabla}\times\vec{B}=\mu_{0}\vec{J},\quad \vec{\nabla}\cdot\vec{B}=0, \quad
\vec{v}\times\vec{B}=\vec{\nabla}\Phi ,
\end{gathered}
\end{eqnarray}
with arbitrary incompressible flow implying  $\varrho =\varrho (\psi)$, and anisotropy function constant on magnetic surfaces, $\sigma _d=\sigma _d(\psi)$.
  Then, according to \cite{cheviakov1}, $\{\vec{B}_1,\, \vec{v}_1,\, \varrho _1,\,  p_{\perp _1},\, p_{\parallel _1}  \}$ defined by the following symmetry transformations
 \begin{eqnarray}
 \begin{gathered}
 \label{transfCGL}
  \vec{B}_1=\frac{b(\vec{r})}{n(\vec{r})}\vec{B}+\frac{c(\vec{r})\sqrt{\mu _0 \varrho}}{n(\vec{r})\sqrt{1-\sigma _d}}\vec{v}, \quad
 \vec{v}_1=\frac{c(\vec{r})\sqrt{1-\sigma _d}}{a(\vec{r})\sqrt{\mu _0 \varrho}}\vec{B}+\frac{b(\vec{r})}{a(\vec{r})}\vec{v}\\
 \varrho _1 (\vec{r})=a^2(\vec{r})\varrho ,\quad
 p_{\perp 1}=C\left(p_{\perp}+\frac{B^2}{2\mu _0}\right)-\frac{B_1^2}{2\mu _0}, \\
 p_{\parallel 1}=C\left(p_{\perp}+\frac{B^2}{2\mu _0}\right)+\left[1-2n^2(\vec{r})(1-\sigma _d)\right]\frac{B_1^2}{2\mu _0}, \\
 C\equiv b^2(\vec{r})-c^2(\vec{r})=\mbox{const.}\neq 0,
 \end{gathered}
 \end{eqnarray}
 is also a solution.
 Note that transformations (\ref{transfCGL}) depend on the arbitrary functions $a(\vec{r})$, $b(\vec{r})$,
  $c(\vec{r})$, $n(\vec{r})$, and reduce to the respective ones for isotropic pressure given by the set (\ref{transfMHD}) when $\sigma_d =0$  and $n(\vec{r})=1$.
\par As stated in \cite{cheviakov1} the functions $a(\vec{r})$, $b(\vec{r})$,
  $c(\vec{r})$, $n(\vec{r})$ have to be constant on the magnetic surfaces. Below we examine the validity of these transformations and whether they can break the geometrical symmetry of the original equilibria.
\subsubsection{\label{4.1.1}Validation of equilibrium equations for the transformed fields}
In order for the new solution (\ref{transfCGL}) to be valid it must satisfy the following set of CGL equilibrium equations
\begin{eqnarray}
\begin{gathered}
\label{CGLnew}
\vec{\nabla}\cdot (\varrho _1\vec{v}_1)=0, \\
\varrho _1(\vec{v}_1\cdot\vec{\nabla})\vec{v}_1=\vec{J}_1\times\vec{B}_1-\vec{\nabla}\cdot{\bf P}_1, \\
\vec{\nabla}\times\vec{B}_1=\mu_{0}\vec{J}_1, \\
\vec{\nabla}\times\vec{E}_1=0\Longrightarrow \vec{E}_1=-\vec{\nabla}\Phi _1 , \\
\vec{\nabla}\cdot\vec{B}_1=0, \\
\vec{E}_1+\vec{v}_1\times\vec{B}_1=0,
\end{gathered}
\end{eqnarray}
where ${\bf P}_1$ and $\sigma_{d_1}$ are given by (\ref{tensor}).
\par
Expressing in (\ref{CGLnew}) the transformed fields in terms of the original ones by means of (\ref{transfCGL})  leads to the following system of  equations:
\begin{eqnarray}
\label{s1}
\vec{B}\cdot \vec{\nabla}b+\Lambda \vec{v}\cdot \vec{\nabla}c-\left(b\vec{B}+c\Lambda \vec{v}\right)\cdot \frac{\vec{\nabla}n}{n}=0, \\
\label{s2}
\Lambda \vec{v}\cdot \vec{\nabla}b+\vec{B}\cdot \vec{\nabla}c+\left(\Lambda b\vec{v}+c\vec{B}\right)\cdot \frac{\vec{\nabla}a}{a}=0, \\
\label{s3}
\vec{B}\cdot \left(\frac{\vec{\nabla}a}{a}+\frac{\vec{\nabla}n}{n}\right)=0, \\
\label{s4}
\vec{v}\cdot \left(\frac{\vec{\nabla}a}{a}+\frac{\vec{\nabla}n}{n}\right)=0, \\
\label{s5}
-\vec{B}\cdot \left(b^2\frac{\vec{\nabla}n}{n}+c^2\frac{\vec{\nabla}a}{a}\right)-bc\Lambda \vec{v}\cdot\left(\frac{\vec{\nabla}n}{n}+\frac{\vec{\nabla}a}{a}\right)+\Lambda \vec{v}\cdot\left(b\vec{\nabla}c-c\vec{\nabla}c\right)=0,\\
\label{s6}
\Lambda \vec{v}\cdot \left(c^2\frac{\vec{\nabla}n}{n}+b^2\frac{\vec{\nabla}a}{a}\right)+bc\vec{B}\cdot\left(\frac{\vec{\nabla}n}{n}+\frac{\vec{\nabla}a}{a}\right)+ \vec{B}\cdot\left(b\vec{\nabla}c-c\vec{\nabla}c\right)=0,
\end{eqnarray}
where $\Lambda \equiv \sqrt{\mu _0 \varrho}/{\sqrt{1-\sigma _d}}$.
\par It is apparent  that if all four functions appearing in the symmetry transformations are constant on the magnetic surfaces, Eqs.  (\ref{s1})-(\ref{s6}) are trivially satisfied; otherwise the above system of six equations for the four functions $a(\vec{r})$, $b(\vec{r})$,
 $c(\vec{r})$, $n(\vec{r})$  is in general  overdetermined. However,
if the functions $a(\vec{r})$, $b(\vec{r})$,
 $c(\vec{r})$, $n(\vec{r})$ are chosen so that
 \begin{equation}
 -\frac{\vec{\nabla}a}{a}=\frac{\vec{\nabla}n}{n}=\frac{\vec{\nabla}(b+c)}{(b+c)},
 \end{equation}
being satisfied when 
 \begin{equation}
  \label{specialchoice}
  a=\frac{1}{n}, \quad n=b+c,
  \end{equation}
then  (\ref{s3}) and  (\ref{s4}) are trivially satisfied, while  (\ref{s1}), (\ref{s2}), (\ref{s5}) and (\ref{s6}) reduce to the  single relationship:
\begin{equation}
\label{reduce}
(\vec{B}-\Lambda \vec{v})\cdot \left(b\vec{\nabla}c-c\vec{\nabla}b\right)=0.
\end{equation}
Since $b\neq \pm c$ for the transformation to be invertible, Eq. (\ref{reduce}) is satisfied only for parallel flows:
\begin{equation}
\label{paral2}
\vec{v}=\frac{\sqrt{1-\sigma _d}}{\sqrt{\mu _0 \varrho}} \vec{B}.
\end{equation}
Note that the field-aligned equilibria (\ref{paral2}) and (\ref{fieldaligned1}) differ from each other, and thus, transformations (\ref{transfCGL}) introduced in \cite{cheviakov1} for flow of arbitrary direction are not reducible into the respective transformations (\ref{transfCGLpar}) for parallel flows presented herein (cf Section II). This result also holds for the respective isotropic transformations (\ref{transfMHD}) and (\ref{transfMHDpar}) derived in Ref. \cite{bogo1}.
 For the  special equilibria  with field-aligned flows satisfying  (\ref{paral2}) and $a(\vec{r})$, $b(\vec{r})$,
  $c(\vec{r})$, $n(\vec{r})$ generally not constant on magnetic surfaces, transformations (\ref{transfCGL}) reduce to 
 \begin{eqnarray}
 \begin{gathered}
 \label{aniso}
 \vec{B}_1=\vec{B}, \quad
 \vec{v}_1=(b+c)^2\vec{v}, \quad
 \varrho _1=\varrho /(b+c)^2, \\
 p_{\parallel _1}=Cp_{\perp}+\left[C+1-2(b+c)^2(1-\sigma _d)\right]\frac{B^2}{2\mu _0}, \\
 p_{\perp _1}=Cp_{\perp}+\left(C-1\right)\frac{B^2}{2\mu _0}, \quad   \sigma _{d_1}=1-(b+c)^2(1-\sigma _d),\\
 C=b^2-c^2=\mbox{const.} \neq 0.
\end{gathered}
 \end{eqnarray}

 For equilibria with isotropic pressure satisfying (\ref{transfMHD}), being recovered from (\ref{transfCGL})  for $\sigma _d=0$ and  $ n=a=b+c=1$,  the choice (\ref{specialchoice})  leads to
  \begin{eqnarray}
  \begin{gathered}
  \label{iso}
  \vec{B}_1=\vec{B},\quad
  \vec{v}_1=\vec{v},\quad
  \varrho _1=\varrho , \\
  p_1=Cp+\left(C-1\right)\frac{B^2}{2\mu _0}, \\
  C=b-c=\mbox{const.} \neq 0.
  \end{gathered}
  \end{eqnarray}
 With the aid of (\ref{aniso}) and (\ref{iso}) we observe that in the presence of pressure anisotropy the transformed velocity and mass density differ from the respective, original ones.
  \par Now suppose that the original equilibrium  is helically symmetric \cite{evangelias2}. Then the following relations hold
  \begin{eqnarray}
  \vec{B}=I\vec{h}+\vec{h}\times \vec{\nabla}\psi , \\
  \vec{v}=\frac{\Theta}{\varrho}\vec{h}+\frac{M_p}{\sqrt{\mu _0 \varrho}}\vec{h}\times \vec{\nabla}\psi ,\\
 \frac{1}{q} \frac{d\Phi }{d\psi}=\frac{I M_p}{\sqrt{\mu _0 \varrho}}-\frac{\Theta}{\varrho},
  \end{eqnarray}
  where the function $\Theta$ relates to the helical velocity field and $\Phi=\Phi(\psi)$.
 Equation (\ref{specialchoice}) implies $I=(\sqrt{1-\sigma _d}/\sqrt{\mu_0 \varrho})(\Theta/\varrho)$  and $d\Phi/d\psi=0$. These relations lead to the following one
 \begin{equation}
 \frac{I}{\sqrt{\mu_0 \varrho}}\left(M_p-\sqrt{1-\sigma _d}\right)=0,
 \end{equation}
 which implies either $I=0$ or $M_p^2+\sigma _d=1$.
 It turns out again that symmetry breaking is possible only for purely poloidal parallel flows $(I=0)$. The relation $M_p^2+\sigma_d=1$ is connected to the Alfv\'en singularity. The same conclusion  holds for axially and translationally symmetric original equilibria with or without pressure anisotropy.
 \subsubsection{\label{4.1.2}Arbitrary functions constant on magnetic surfaces}
 In the above Subsection we found that the symmetry transformations (\ref{transfCGL}) (and the respective transformations (\ref{transfMHD}) for isotropic pressure) are valid when the arbitrary functions are constant on magnetic surfaces, since Eqs.  (\ref{s1})-(\ref{s6}) are trivially satisfied. Here we examine the equilibria derived from a given geometrically symmetric one of this kind.
 \par Let the original CGL equilibria (\ref{CGL}) posses magnetic surfaces $\psi =$const.  which both $\vec{B}$ and $\vec{v}$ lie on. Also, suppose that respective surfaces $\psi _1=$const. are defined for the transformed equilibria (\ref{transfCGL})  which $\vec{B}_1$ and $\vec{v}_1$ lie on. It holds that
 \begin{equation}
  \vec{B}_1\times \vec{v}_1=\frac{C}{n(\psi)a(\psi)}\vec{B}\times \vec{v},
 \end{equation}
 and thus, the magnetic surfaces through the transformation are preserved:
 \begin{equation}
  \psi _1=F(\psi).
 \end{equation}
 This means that all vectors $\vec{B},\, \vec{B}_1,\, \vec{v},\, \vec{v}_1$ lie on the surfaces $\psi =$const. As a result, if the original equilibria has some known geometrical symmetry, the transformed equilibria will have the same symmetry, too.
 \newline
 \newline
 Consider now helically symmetric equilibria with incompressible flow of arbitrary direction and anisotropy function constant on magnetic surfaces \cite{evangelias2}:
 \begin{eqnarray}
 \begin{gathered}
 \label{helsymm}
 \vec{B}=I\vec{h}+\vec{h}\times \vec{\nabla}\psi(r,u) , \\
 \vec{v}=\frac{\Theta}{\varrho}\vec{h}+\frac{M_p}{\sqrt{\mu _0 \varrho}}\vec{h}\times \vec{\nabla}\psi(r,u), \\
 \mu _0\vec{J}=(\mathcal{L}\psi (r,u)+2kmqI(\psi ,r))\vec{h}-\vec{h}\times \vec{\nabla}I(\psi ,r), \\
 \bar{p}=\bar{p}_s(\psi)-\varrho \left[\frac{v^2}{2}-\frac{(1-\sigma _d)}{q(1-\sigma _d-M_p^2)}\left(\frac{d\Phi}{d\psi}\right)^2\right],
 \end{gathered}
 \end{eqnarray}
 where the elliptic operator is defined as $\mathcal{L}:=\left(\vec{\nabla}\cdot (q\vec{\nabla})\right)/q$. Note that the current density lies on well defined helicoidal surfaces $I=$const., while the effective pressure is uniform on the surfaces defined by $\bar{p}=$const., both of these two sets of surfaces not coinciding with the magnetic surfaces.
 By applying the symmetry transformations (\ref{transfCGL}) with $a=a(\psi), \,b=b(\psi), \,c=c(\psi), \,n=n(\psi)$, we obtain the following class of equilibria:
 \begin{eqnarray}
 \vec{B}_1=
 \underbrace{\frac{1}{n}\left(bI+c\frac{\sqrt{\mu _0 \varrho}}{\sqrt{1-\sigma _d}} \frac{\Theta}{\varrho}\right)}_
 {I_1} \vec{h}
 +\underbrace{\frac{1}{n}\left(b+c\frac{M_p}{\sqrt{1-\sigma _d}}\right)}_{\text{G}}\vec{h}\times \vec{\nabla} \psi ,
 \\
\label{vtransf}
 \vec{v}_1=\frac{1}{a}\left(c\frac{\sqrt{1-\sigma _d}}{\sqrt{\mu _0 \varrho}}I+b\frac{\Theta}{\varrho}\right)\vec{h}+\frac{1}{a\sqrt{\mu _0\varrho}}\left(c\frac{\sqrt{1-\sigma _d}}{\sqrt{\mu _0 \varrho}}+b\frac{M_p}{\sqrt{1-\sigma _d}}\right)\vec{h}\times \vec{\nabla} \psi ,\\
 \mu _0 \vec{J}_1=\left(G\mathcal{L}\psi +\frac{dG}{d\psi}|\vec{\nabla}\psi |^2+2kmqI_1\right)\vec{h}-\vec{h}\times \vec{\nabla}I_1 ,\\ 
 \bar{p}_1=C\bar{p}+(1-\sigma _d)\frac{(CB^2-n^2B_1^2)}{2\mu _0}.
 \end{eqnarray}
 	Note that although the magnetic surfaces are preserved, neither the transformed current density nor the transformed effective pressure remain on the surfaces of the respective original quantities: $\vec{J}_1\cdot \vec{\nabla}I\neq 0$, $\bar{p}_1\neq \bar{p}$.
 \par
 Now since $\psi _1=F(\psi)$ and  the  original equilibria are helically symmetric the  transformed ones  should retain that  symmetry. This means that the transformed fields can also be written in a form similar to (\ref{helsymm});  in particular for the transformed velocity we have:
 \begin{equation}
 \label{vsymm}
 \vec{v}_1=\frac{\Theta _1}{\varrho _1}\vec{h}+\frac{M_{p _1}}{\sqrt{\mu _0 \varrho _1}}\vec{h}\times \vec{\nabla}\psi _1(r,u),
 \end{equation}
 where
 \begin{equation}
  M_{p_1}^2=\frac{v_{pol _1}^2}{B_{pol _1}^2/\mu _0\varrho _1}=(n\sqrt{1-\sigma _d})\frac{c\sqrt{1-\sigma _d}+bM_p}{b\sqrt{1-\sigma _d}+cM_p}.
  \end{equation}
 Equality of  the poloidal velocity components in  (\ref{vtransf}) and (\ref{vsymm}) yields
 \begin{equation}
 \label{psipsi1}
 \frac{d\psi _1}{d\psi}=\frac{b\sqrt{1-\sigma _d}+cM_p}{n\sqrt{1-\sigma _d}}\Rightarrow \psi _1(\psi)=\int_{0}^{\psi} \frac{b(\chi)\sqrt{1-\sigma _d(\chi)}+c(\chi)M_p(\chi)}{n(\chi)\sqrt{1-\sigma _d(\chi)}}d\chi .
 \end{equation}
As already mentioned in Subsection \ref{2.2} every equilibrium  static or stationary with incompressible flows, which has some geometrical symmetry 
 with pressure anisotropy function constant on magnetic surfaces, is governed by a GS equation for the flux function $\psi$  \cite{clemente,evangelias1,evangelias2,throumtasso,agim,simintzis}. Such an equation  contains a quadratic term as $|\vec{\nabla}\psi|^2$. For this reason an integral  transformation is applied as:
 \begin{equation}
 \label{transfgen}
 U(\psi)=\int_{0}^{\psi} \sqrt{1-\sigma _{d }(\chi)-M_{p}^2(\chi)}d\chi .
 \end{equation}
 Transformation (\ref{transfgen}) does not affect the magnetic surfaces, it just relabels them by the flux function $U$. Furthermore, the respective GS equilibrium equations can be solved by analytical techniques in the $U$-space, since (\ref{transfgen})  eliminates a quadratic term as $|\vec{\nabla} U|^2$. Transformation (\ref{transfgen}) introduced in \cite{evangelias1, evangelias2}, is a generalisation of that introduced in \cite{simintzis} for isotropic equilibria with incompressible flow ($\sigma _d=0$) and that introduced in \cite{clemente} for static anisotropic equilibria ($M_p^2=0$).
 \par Adopting (\ref{transfgen}) both for the original and the transformed helically symmetric equilibria,  Eq. (\ref{psipsi1}) yields
 \begin{equation}
  \frac{dU_1(U)}{dU}=C^{1/2} \Rightarrow U_1=C^{1/2}U.
  \end{equation}
  Therefore the transformed equilibria differ from the starting ones only by a constant factor $C^{1/2}$, in agreement with the conclusions drawn  in the previous Sections; the geometrical symmetry of the original equilibria can break only for purely poloidal magnetic field, otherwise the transformed equilibria retains the original symmetry.
 \subsection{\label{4.2}Transformations between MHD-CGL equilibria}
 In Refs. \cite{cheviakov1,cheviakov3,cheviakov2} transformations that produce CGL anisotropic equilibria from given isotropic MHD ones, are introduced as follows:
 If $\{ \vec{B},\, \vec{v},\, p,\, \varrho \}$ is a known solution of the MHD equilibrium system (\ref{MHD}), then the following symmetry transformations
 \begin{eqnarray}
 \begin{gathered}
 \label{mixed}
 \vec{B}_1=f_1(\vec{r})\vec{B} ,\quad
 \vec{v}_1=g_1(\vec{r})\vec{v} ,\quad
 \varrho _1=\frac{C_0 \mu _0}{g_1^2(\vec{r})}\varrho , \\
 p_{\perp _1}=C_0 \mu _0 p+C_1+\left(C_0\mu _0 -f_1^2(\vec{r})\right)\frac{B^2}{2\mu _0}
, \\
p_{\parallel _1}=C_0 \mu _0 p+C_1-\left(C_0\mu _0 -f_1^2(\vec{r})\right)\frac{B^2}{2\mu _0} ,
\end{gathered}
\end{eqnarray}
where $C_0$ and $C_1$ are arbitrary constants, produce an infinite family of CGL equilibria satisfying (\ref{CGLnew}). Transformations (\ref{mixed}) are also valid in the static limit, $\vec{v}=0$. Let us examine their validity.
\par Substituting (\ref{mixed}) into (\ref{CGLnew}) we obtain 
\begin{eqnarray}
\begin{gathered}
\vec{B}\cdot \vec{\nabla}f_1(\vec{r})=0, \\
\vec{v}\cdot \vec{\nabla}g_1(\vec{r})=0.
\end{gathered}
\end{eqnarray}
Thus, in order for transformations (\ref{mixed}) to be valid, the functions $f_1(\vec{r})$ and $g_1(\vec{r})$ must be constant on the magnetic field lines and velocity  streamlines of the original equilibria, and respective considerations on their structure can be made as those in Section \ref{2}. Therefore it turns out again that the only way that the geometrical symmetry of the original isotropic equilibria can break is if and only if the magnetic and velocity fields are collinear and purely poloidal. In this context, conclusions for the breaking of the helical symmetry of astrophysical jets with magnetic field lines going to  infinity in connection with the coordinate  $z$, examined in \cite{cheviakov1,cheviakov2} in the static limit, should be reconsidered.
 
 \section{\label{5}Conclusions}
 In the present work we made an extensive revision of the symmetry transformations previously introduced in a series of papers, \cite{bogo1,bogo,bogo3,bogo2,cheviakov1,cheviakov3,cheviakov2,anco}, which once applied to known MHD and/or CGL equilibria produce an infinite new continuous families of respective equilibria. These transformations contain some arbitrary scalar  functions, the structure of which depend on the topology of the given equilibria. We examined both transformations that map MHD into MHD, CGL into CGL, and MHD into CGL equilibria, either with field-aligned or arbitrary incompressible flows, particularly as concerns their validitity and applicability. In addition, we examined whether these transformations can break the geometrical symmetry of the original equilibria.
 \par In Section \ref{2} we presented a new set of symmetry transformations that can be applied to any known CGL equilibria with special field-aligned incompressible flow satisfying (\ref{fieldaligned1}) and pressure anisotropy function, $\sigma_d$, constant on the magnetic field lines, to produce an infinite class of equilibria with collinear $\vec{v}_1$ and  $\vec{B}_1$ fields, and with $\varrho$ and $\sigma _d$ functions that in general may be arbitrary. These transformations consist a generalisation of the ones introduced in \cite{bogo1} for the same kind of field-aligned incompressible flow and isotropic pressure, and can also be applied to static anisotropic equilibria.
 \par In addition, we examined the structure of the arbitrary scalar functions included in the symmetry transformations in relation with the topology of the magnetic field of the original equilibrium and the existence of magnetic surfaces, and proved that if  the original equilibrium possesses some known geometrical symmetry, this can be broken by the transformations if and only if the magnetic field is purely poloidal. In this respect, in Section \ref{3} we applied the aforementioned symmetry transformations to specifically prescribed  axisymmetric CGL equilibria with collinear and purely poloidal $\vec{v}$ and  $\vec{B}$ fields, incompressible flow, and $\sigma_d$ uniform on the magnetic surfaces, and constructed 3D equilibria with mass density and anisotropy function that may vary on the magnetic surfaces.
 \par In Section \ref{4} we examined the transformations introduced in \cite{bogo1,bogo,bogo3,bogo2,cheviakov1,cheviakov3,cheviakov2,anco} applied to given equilibria with incompressible flow non-collinear to the magnetic field. We showed that these transformations are valid if the arbitrary functions included therein are either constant on the magnetic surfaces, if such surfaces exist, or if they are  related by a special relationship; in the latter case it turns out that   the fields $\vec{v}$ and  $\vec{B}$ of the original equilibria are restricted to be collinear. If the original equilibria have certain  geometrical symmetry, in the former case they differ from the trasformed ones only by a constant factor, while in the latter case this symmetry can break only for purely poloidal magnetic fields.
 \par Summarizing, the generic conclusion of this study is that both transformations introduced in \cite{bogo1,bogo,bogo3,bogo2,cheviakov1,cheviakov3,cheviakov2,anco} and the ones presented herein (cf Section \ref{2}) can break the geometrical symmetry of the original equilibria, both static and/or with field-aligned incompressible flow and both isotropic and/or anisotropic with the function $\sigma_d$ constant on the magnetic field  lines, if and only if the magnetic field is purely poloidal. Otherwise the transformed equilibria retain the geometrical symmetry of the original ones.
\begin{acknowledgments}
 The authors would like to thank Prof. Oleg I. Bogoyavlenskij for useful discussion related with the symmetry breaking of the axisymmetric Hill's vortex equilibrium.
 \par
 This study was performed within the framework of the EUROfusion Consortium and has received funding from the National Program for the Controlled Thermonuclear Fusion, Hellenic Republic. The views and opinions expressed herein do not necessarily reflect those of the European Commission. A.E. has
been financially supported by the General Secretariat for Research and
Technology (GSRT) and the Hellenic Foundation for Research and
Innovation (HFRI).
 \end{acknowledgments}


\begin{thebibliography}{9}
 
 \bibitem{CGL}
 G. F. Chew, M. L. Goldberger, and F. E. Low, Proc. R. Soc. A \textbf{236}, 112 (1956).
 
\bibitem{bogo1} 
 O. I. Bogoyavlenskij, Phys. Rev. E \textbf{66}, 056410 (2002).
 
 \bibitem{bogo}
 O. I. Bogoyavlenskij, Phys. Lett. A \textbf{291}, 256 (2001).
 
 \bibitem{bogo3}
 O. I. Bogoyavlenskij, Phys. Lett. A \textbf{276}, 257 (2000).
 
\bibitem{bogo2}
 O. I. Bogoyavlenskij, Phys. Rev. E \textbf{62}, 8616 (2000) . 
 
 \bibitem{cheviakov1} 
 A. F. Cheviakov and O. I. Bogoyavlenskij, J. Phys. A: Math Gen. \textbf{37}, 7593 (2004).
 
 \bibitem{cheviakov3}
 A. F. Cheviakov, Phys. Rev. Lett. \textbf{94}, 165001 (2005).
 
\bibitem{cheviakov2} 
 A. F. Cheviakov, Topology and its Applications \textbf{152}, 157 (2005).
 
 \bibitem{anco}
 A. F. Cheviakov and S. C. Anco, Phys. Lett. A \textbf{372}, 1363 (2008).
 
 \bibitem{kruskal}
  M. D. Kruskal and R. M. Kulsrud, Phys. Fluids \textbf{1}, 265 (1958).
 
 \bibitem{hopf} P. Alexandroff and H. Hopf, Topologie (Springer, Berlin, 1935) p. 552.

 
 \bibitem{morozov}
  A. I. Morozov and L. S. Solov'ev, Soviet Phys. JETP \textbf{18}, 660 (1964).
 
 \bibitem{grad}
 H. Grad, Phys. Fluids \textbf{10}, 137 (1967).
 
 \bibitem{moffatt}
 H. K. Moffatt, On the existence, structure and stability of MHD equilibrium states, in: Turbulence and Nonlinear Dynamics in MHD Flows, edited by: M. Meneguzzi, A. Pouquet, and P. L. Sulem (North-Holland, Amsterdam, 1989) p. 185.
 
 \bibitem{moawad}
 S. M. Moawad and D .A. Ibrahim, Phys. Plasmas \textbf{23}, 082502 (2016).
 
 \bibitem{shafranov}
 V. D. Shafranov, Soviet Phys. JETP \textbf{6}, 545 (1958).
 
 \bibitem{hill}
 J. Hill, Philos. Trans. R. Soc. London, Ser. A \textbf{185} 213 (1894).
 
 \bibitem{johnson}
 J. L. Johnson, C. R. Oberman, R. M. Kulsrud, and E. A. Frieman, Phys. Fluids \textbf{1}, 281 (1958).
 
 \bibitem{bogo4}
 O. I. Bogoyavlenskij, Phys. Rev. Lett. \textbf{84}, 1914 (2000).
 
 \bibitem{clemente}
 R. A. Clemente, Nucl. Fusion \textbf{33}, 963 (1993).
 
\bibitem{evangelias1} 
 A. Evangelias and G. N. Throumoulopoulos, Plasma Phys. Control. Fusion \textbf{58}, 045022 (2016).
 
 \bibitem{evangelias2}
 A. Evangelias, A. Kuiroukidis, and G. N. Throumoulopoulos, Plasma Phys. Control. Fusion \textbf{60}, 025005 (2018).
 
 
 \bibitem{mercier}
 C. Mercier and M. Cotsaftis, Nucl. Fusion \textbf{1}, 121 (1961).
 
  \bibitem{throumtasso}
  G. N. Throumoulopoulos and H. Tasso, Phys. Plasmas \textbf{4}, 1492 (1997)
 
 \bibitem{agim}
  Y. Z. Agim and J. A. Tataronis, J. Plasma Physics \textbf{34}, 337 (1985).

\bibitem{simintzis}
C. Simintzis, G. N. Throumoulopoulos, G. Pantis, and H. Tasso, Phys. Plasmas \textbf{8}, 2641 (2001).

\end{thebibliography}
 \end{document}